\documentclass[prl,twocolumn,superscriptaddress,showpacs,preprintnumbers,amsmath,amssymb]{revtex4}
\usepackage{amssymb}

\usepackage{graphicx}
\usepackage{dcolumn}
\usepackage{bm}
\usepackage{epsfig}
\newcommand{\ignore}[1]{}

\begin{document}

\title{Quantum Stress: Density Functional Theory Formulation and Physical Manifestation }


\author{Hao Hu}
\affiliation{Department of Materials Science and Engineering
University of Utah, Salt Lake City, UT 84112, USA}

\author{Zhengfei Wang}
\affiliation{Department of Materials Science and Engineering
University of Utah, Salt Lake City, UT 84112, USA}

\author{Junyi Zhu}
\thanks{Present address:  National renewable energy Lab, 1617 Cole Blvd. Golden, CO 80401}
\affiliation{Department of Materials Science and Engineering
University of Utah, Salt Lake City, UT 84112, USA}

\author{Dangxin Wu}
\affiliation{Department of Materials Science and Engineering
University of Utah, Salt Lake City, UT 84112, USA}

\author{Miao Liu}
\affiliation{Department of Materials Science and Engineering
University of Utah, Salt Lake City, UT 84112, USA}

\author{Hepeng Ding}
\affiliation{Department of Materials Science and Engineering
University of Utah, Salt Lake City, UT 84112, USA}

\author{Zheng Liu}
\affiliation{Department of Materials Science and Engineering
University of Utah, Salt Lake City, UT 84112, USA}
\affiliation{Insitute for Advanced Study, Tsinghua University, Beijing, China, 100084}

\author{Feng Liu}
\thanks{Corresponding author: fliu@eng.utah.edu}
\affiliation{Department of Materials Science and Engineering
University of Utah, Salt Lake City, UT 84112, USA}

\begin{abstract}
The concept of "quantum stress (QS)" is introduced and formulated within density functional theory (DFT), to elucidate extrinsic electronic effects on the stress state of solids and thin films in the absence of lattice strain. A formal expression of QS ($\sigma^Q$) is derived in relation to deformation potential of electronic states ($\Xi$) and variation of electron density ($\Delta n$), $\sigma^Q=\Xi \Delta n$, as a quantum analog of classical Hook's law. Two distinct QS manifestations are demonstrated quantitatively by DFT calculations: (1) in the form of bulk stress induced by charge carriers; and (2) in the form of surface stress induced by quantum confinement. Implications of QS in some physical phenomena are discussed to underlie its importance.

\end{abstract}

\maketitle
A fundamental property of solid materials is their stress state. At the equilibrium lattice constant, the bulk of a crystalline solid is stress free, but the surface of a solid has intrinsic non-zero stress and stress is commonly induced by any form of lattice distortion \cite{ref1}. The stress (strain) state of a solid or thin-film material has profound effects on its thermodynamic stability and physical and chemical properties \cite{ref1,ref2,ref3}, and has been employed in a wide range of applications such as electromechnical devices \cite{ref4}, mechanochemcial sensors \cite{ref5} and flexible electronics \cite{ref6}, and even to make new nanotructures \cite{ref7,ref8}. Here, we introduce the concept of "quantum stress (QS)", which adds an interesting quantum mechanical aspect to our conventional view of "classical stress (CS)" based on classical mechanics. We mathematically formulate the expression of QS within density function theory (DFT), and use DFT calculations to demonstrate quantitatively two distinct physical manifestations of QS, in the form of bulk stress induced by charge carriers in a homogeneous system of crystalline solids and in the form of surface stress induced by quantum confinement in a heterogeneous system of nanoscale thin films. We will then apply the concept of QS to elucidate a few examples of physical phenomena that underlie the importance and usefulness of QS.

\begin{figure}[b]
\includegraphics[width=3.4in]{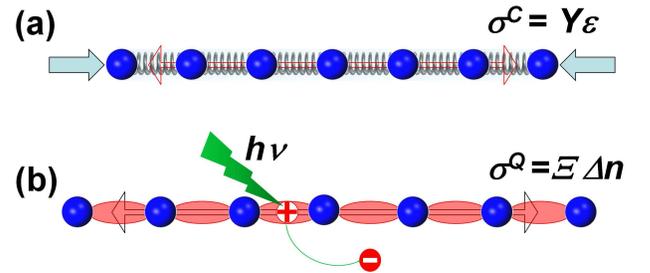}
\caption{(Color online) Schematic illustration of CS versus QS. (a) The CS ($\sigma^C$) induced by applying a compressive lattice strain ($\varepsilon $). Arrows indicate stress and force directions. (b) The QS ($\sigma^Q$) induced by a hole excited by a photon. \label{fig1}}
\end{figure}
\emph{Concept of QS}. We first introduce the concept of QS in contrast with the CS. Figure 1 illustrates the fundamental difference between the QS and CS using a simple model of a one-dimensional (1D) lattice. Consider a lattice is under compressive (Fig. 1a) or tensile lattice strain ($\varepsilon$), such as in an epitaxial film due to lattice mismatch between the film and substrate \cite{ref9,ref10}. The "atomic" lattice deformation energy can be expressed as $E = (1/2)Y\varepsilon^2 V$, where $Y$ is Young's modulus and $V$ is the volume of lattice. By definition, the lattice formation induced lattice stress, which we refer to here as CS, is expressed as $\sigma^C = (1/V)(dE/d\varepsilon) = Y\varepsilon$ , the Hook's law. Now, consider an equilibrium lattice in the absence of strain ($\varepsilon  = 0$), but electronically perturbed or excited, such as an electron is kicked out by a photon leaving behind a hole, as shown in Fig. 1b, which redistributes the electron density. The "electronic deformation" energy can be expressed as $E = \mu \Delta N$, where $\mu$  is electron chemical potential and  $\Delta N$ is the change of number of electrons. Then, the lattice stress induced by the electronic deformation, which we refer to as QS, can be expressed as $\sigma^Q = (1/V)(dE/d\varepsilon ) = \Xi \Delta n$, where  $\Xi=d\mu/d\varepsilon$  is deformation potential and $\Delta n$ is change of electron density. The expression of $\sigma^Q = \Xi \Delta n$ can be viewed as a quantum analog of Hook's law. Below, we provide a formal derivation of CS and QS within DFT.

\emph{DFT formulation}. Following DFT \cite{ref11}, the total energy functional of a solid is written as
\begin{equation}
\label{Eq_1}
 E[n(\vec{r}),\{\vec{R}_m\}]=E_e[n(\vec{r})]+E_{ext}[n(\vec{r}),\{\vec{R}_m\}]+E_{I}[\{\vec{R}_m\}]
\end{equation}

$E_e[n(\vec{r})]$is the electronic energy functional of charge density $n(\vec{r}$, including kinetic and electron-electron interaction energy, $E_{ext}[n(\vec{r}),\{\vec{R}_m\}]$  is the ion-electron interaction energy, $E_{I}[\{\vec{R}_m\}]$ is the ion-ion interaction energy and $\{\vec{R}_m\}$  are atomic coordinates.  First, for completeness, let's rederive the Hook's law, the expression of CS under finite lattice strain. Consider coordinate transformation $\vec{r}=(1+\{{\varepsilon_{ij}}\})\vec{r}^0$  and ${\vec{R}_m}=(1+\{\varepsilon_{ij}\}){\vec{R}^0_m}$  under strain $\{\varepsilon_{ij}\}$ , where $\vec{r}^0$ and $\{\vec{R}^0_m\}$ are the electronic and atomic coordinates of strain-free equilibrium lattice. Let $n^0(\vec{r}^0)$ and $n^{\varepsilon}(\vec{r})$  be the ground-state charge density before and after strain is applied. By definition, the stress tensor is expressed as
\begin{equation}
\label{Eq_2}
\begin{split}
 &\sigma^C_{ij}=\frac{1}{V}\frac{dE[n(\vec{r}),\{\vec{R}_m\}]}{d\varepsilon_{ij}}|_{n^{\varepsilon},\{\vec{R}_m\}}\\
 &=\frac{1}{V}\left[\int_V \left(\frac{\delta(E_e+E_{ext})}{\delta n(\vec{r})} \right) \frac{\delta n(\vec{r})}{\delta \varepsilon_{ij}}d\vec{r} + \sum_m \frac{\partial E_R}{\partial \vec{R}_m}\frac{\vec{R}_m}{\partial \varepsilon_{ij}} \right]
\end{split}
\end{equation}

where $E_R=E_{ext}+E_{I}$. Since $n^{\varepsilon}(\vec{r})$ is the ground-state electron density at $\vec{r}$  and $\{\vec{R}_m\}$ , according to Hohenberg-Kohn theorem \cite{ref11}, we have $\left(\frac{\delta(E_e+E_{ext})}{\delta n(\vec{r})} \right)_{{n^{\varepsilon},\{\vec{R}_m\}}}=0$  and Eq. (\ref{Eq_2}) becomes
\begin{equation}
\label{Eq_3}
 \sigma^C_{ij}=\frac{1}{V}\left[\frac{\partial E_R}{\partial \varepsilon_{ij}} \right]_{{n^{\varepsilon},\{\vec{R}_m\}}}
\end{equation}
For simplicity, assuming hydrostatic strain,$\varepsilon_{ij}=\varepsilon \delta_{ij}$, we expand $E_R$ in $\varepsilon$
\begin{equation}
\label{Eq_4}
\begin{split}
 E_R[{n^{\varepsilon},\{\vec{R}_m\}}]&=E_R[{n^{\varepsilon},\{\vec{R}^0_m\}}]+\varepsilon \sum_m \vec{R}^0_m \left(\frac{\partial E_R}{\partial \vec{R}_m}\right)_{\vec{R}^0_m}\\
 &+\frac{\varepsilon^2}{2} \sum_m \left[ \left(\vec{R}^0_m \cdot \frac{\partial}{\partial \vec{R}_m} \right)^2 E_R \right]_{\vec{R}^0_m}+...
\end{split}
\end{equation}

Then the stress can be expressed in the first order of $\varepsilon$  as
\begin{equation}
\label{Eq_5}
 \sigma^C=Y \varepsilon
\end{equation}
where $Y=\sum_m \left[\left(\vec{R}^0_m \cdot \frac{\partial}{\partial \vec{R}_m}\right)^2 E_R \right]_{\vec{R}^0_m}$ is the Young's modulus.

Next, we derive the QS induced by electronic excitation and perturbation without applying lattice strain ($\varepsilon_{ij}=0$).  Consider a variation of electron density from $n^0(\vec{r}^0)$ the ground-state density at $\vec{r}^0$ and $\{\vec{R}^0_m\}$  as $n^*(\vec{r}^0)=n^0(\vec{r}^0)+\delta n(\vec{r}^0)$ . (Below, for convenience, we will neglect the superscript 0 for $\vec{r}^0$.) The differentials of energy functionals are
\begin{equation}
\label{Eq_6}
 F[n^*(\vec{r})]=F[n^0(\vec{r})]+\int_V \left( \frac{\delta F[n(\vec{r})]}{\delta n(\vec{r})} \right)_{n^0}\delta n(\vec{r})d\vec{r},
\end{equation}
The stress tensor is
\begin{equation}
\label{Eq_7}
\begin{split}
 \sigma^Q_{ij}&=\frac{1}{V}\frac{dE[n(\vec{r}),\{\vec{R}_m\}]}{d\varepsilon_{ij}}|_{n^*,\varepsilon_{ij}=0}\\
 &=\frac{1}{V}\left\{\int_V \left[\frac{\partial \mu}{\partial \varepsilon_{ij}}\delta n(\vec{r})+\mu \frac{\partial ( \delta n(\vec{r}))}{\partial \varepsilon_{ij}} \right]d\vec{r} \right\}_{n^0,\varepsilon_{ij}=0}
\end{split}
\end{equation}
Where $\mu=\partial^{'}_n(E_e+E_{ext})$  is the electron chemical potential. To arrive at Eq. (\ref{Eq_7}), we used the condition that the strain-free ground-state solid is stress free, i.e., $\left(\frac{dE}{d\varepsilon_{ij}}\right)_{n^0,\varepsilon_{ij}=0}=0$ . It can be shown that the second term in Eq. (\ref{Eq_7}) vanishes because chemical potential remains uniform and the number of electrons is independent of strain, so we have the expression of QS as
\begin{equation}
\label{Eq_8}
 \sigma^Q_{ij}=\frac{1}{V}\left[ \int_V \frac{\partial \mu}{\partial \varepsilon_{ij}}\delta n(\vec{r})d\vec{r} \right]_{n^0,\varepsilon_{ij}=0}
\end{equation}
In a homogeneous crystalline solid, to a good approximation, the electron deformation potential $\Xi=\partial \mu/\partial \varepsilon_{ij}$ is uniform as the electron density remains uniform before and after strain is applied. Then, the expression of QS can be simplified as
\begin{equation}
\label{Eq_9}
 \sigma^Q=\Xi \Delta n
\end{equation}
Equation (\ref{Eq_9}) can be viewed as a quantum analog of Eq. (\ref{Eq_5}), with $\sigma^Q$, $\Xi$  and $\Delta n$ playing the role of $\sigma^C$, $Y$ and $\varepsilon$, respectively. However, Eq. (\ref{Eq_8}) must be used if $\Xi$ is not uniform in a heterogeneous system. For example, in thin films (hetero-junctions) when strain is applied, charge will be redistributed in the surface (interface) regions due to the nonuniform

Notice that the CS reflects the atomic and lattice size effects. Although in the derivation of CS (Eq. (\ref{Eq_5})), the energy expression of $E_R$ has implicitly the electronic contribution through the ion-electron interaction energy $E_{ext}$ as strain changes electron density from $n^0(\vec{r}^0)$ to $n(\vec{r})$ , the stress, nevertheless, follows the Hook's law of classical mechanics depending only on atomic coordinates. In other words, even though the atomic and lattice size is associated with electronic structure, the effects of the ground-state electronic structure can be cast into the atomic and lattice size effect, having a CS manifestation. It is for this reason that the CS can be modeled by empirical interatomic potential involving explicitly only the atomic degrees of freedom as done in molecular dynamics simulations. In contrast, the QS has a pure electronic origin involving explicitly only the electronic degrees of freedom $[n(\vec{r})]$ that cannot be cast into the atomic or lattice size effect. Consequently, the QS must be described by the quantum mechanics of the electronic structure rather than the classical mechanics of the atomic structure. Below, using first-principles quantum-mechanical stress calculations\cite{suppl}, we directly quantify the magnitude and reveal the nature of QS in two distinct manifestations.

\emph{QS induced by charge carrier}. We first demonstrate the QS for the case of a homogeneous system where Eq. (\ref{Eq_9}) can be applied, in the form of bulk stress when an electron is added to or removed from a solid lattice, as in the case of semiconductor doping or photo-excited charge carriers in solids. We have performed DFT calculations \cite{ref12} of the lattice stress induced by adding electrons and/or holes (i.e., removing an electron) to a finite lattice of Al (metal), Si (elemental semiconductor), GaAs (compound semiconductor) and ZrO$_2$ (insulator), and graphite with a hexagonal lattice. Figure 2 shows the calculated $\sigma^Q$ as a function of $\Delta n$ for Al, Si, GaAs and ZrO$_2$, which shows an almost perfect linear dependence for all the cases, in excellent agreement with our theoretical derivation of Eq. (\ref{Eq_9}). In general, electrons induce compressive QS (by convention, compressive stress is defined as negative); while holes induces tensile QS. In plotting Fig. 2, the QS values are taken from the diagonal terms of the stress tensor along the principal axes, since stress is isotropic in a cubic lattice. More generally, electrons or holes may induce anisotropic stress, such as in a hexagonal lattice of graphite (see Fig. 4 below).

\begin{figure}[t]
\includegraphics[width=3.0in]{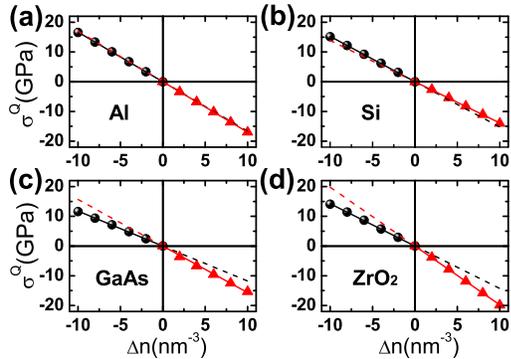}
\caption{(Color online) The QS induced by electrons (triangles) and holes(circles)  as a function of carrier density in (a) Al, (b) Si, (c) GaAs, and (d) ZrO$_2$. Solid lines are linear fits to the data, and the dashed lines are extensions of the solid lines to indicate different slopes for electrons from holes. \label{fig1}}
\end{figure}

According to Eq. (\ref{Eq_9}), the slope of $\sigma^Q$ vs. $\Delta n$ equals to the deformation potential, $\Xi$. For a metal, $\Xi=\frac{\partial E_F}{\partial \varepsilon} $ is the same for electron and hole because of the electron-hole (e-h) symmetry in the metal, as seen for Al in Fig. 2a, and we found $\Xi_{Al}  = -10.49 eV$. For a semiconductor or insulator, however, the deformation potential for electron ($\Xi^e=\frac{\partial E_{CBM}}{\partial \varepsilon}$, CBM stands for conduction band minimum) is different from that for hole ($\Xi^h=\frac{\partial E_{VBM}}{\partial \varepsilon}$, VBM stands for valence band maximum) because of the e-h asymmetry, as seen for Si, GaAs and ZrO$_2$ in Fig. 2. We obtained that $\Xi^e_{Si} =-8.65$, $\Xi^h_{Si} =-9.51$; $\Xi^e_{GaAs} =-9.77$, $\Xi^h_{GaAs} =-7.33$; $\Xi^e_{ZrO_2} =-12.36$, $\Xi^h_{ZrO_2} =-8.87$, which are in good agreement with previous results \cite{ref13}. In general, the larger the band gap is, the larger the e-h asymmetry and hence the larger the difference between the electron and hole deformation potential will be. For Si, $\Xi^e$ is smaller than $\Xi^h$, possibly because it is an indirect semiconductor, while for GaAs and ZrO$_2$, $\Xi^e$ is larger than $\Xi^h$.

We note that conventionally, the deformation potential is derived by calculating the valence and conduction band edge positions as a function of strain, which can be difficult for DFT methods because of the arbitrariness in the absolute value of band energy. Here, our QS calculation provides an efficient and effective method to derive the deformation potential from total energy calculations without the need of calculating band structure.

\begin{figure}[b]
\includegraphics[width=2.8in]{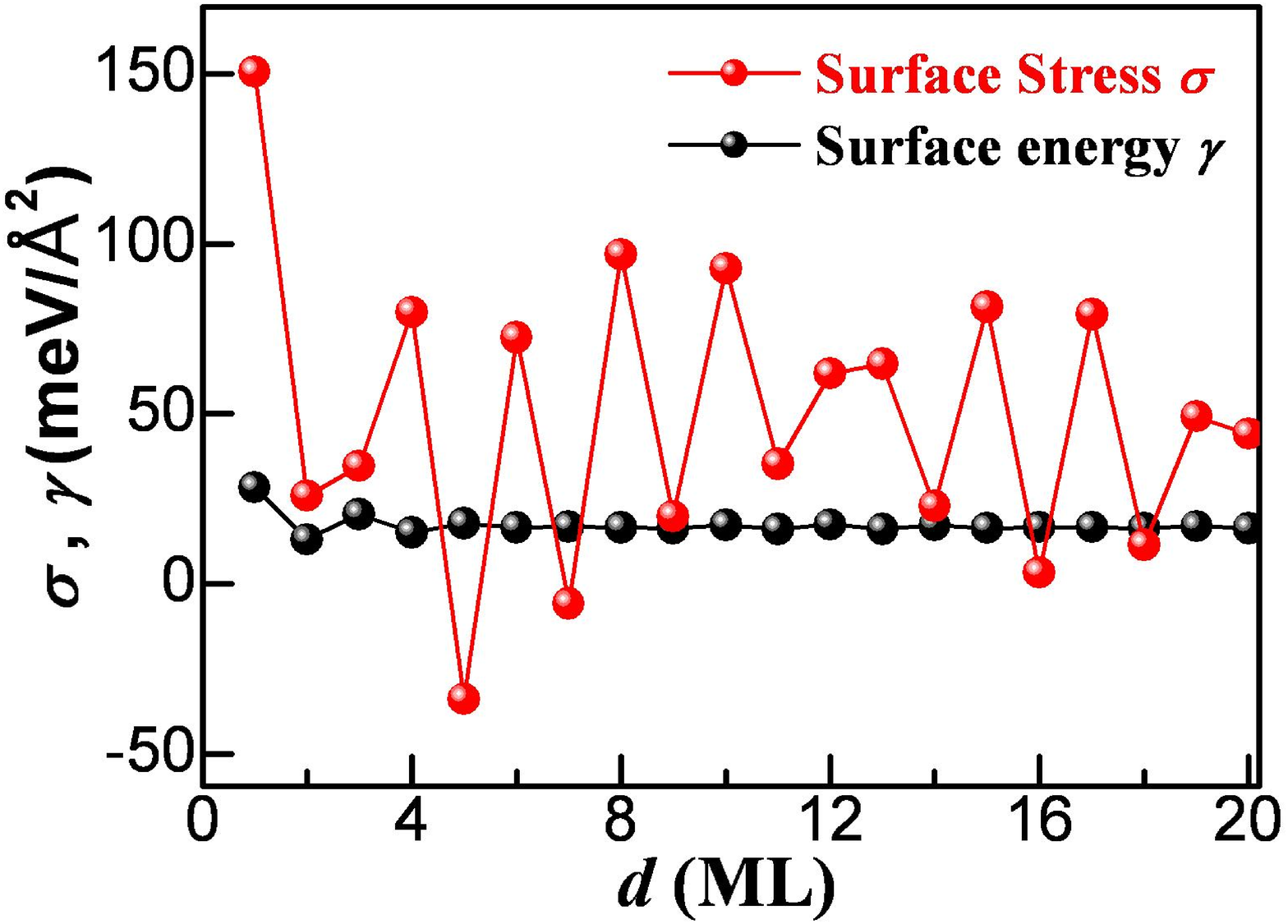}
\caption{(Color online) The calculated surface energy and surface stress of Pb(111) film as a function of film thickness, demonstrating the surface QS in metal nanofilms induced by quantum confinement. \label{fig3}}
\end{figure}

\emph{QS induced by quantum confinement}. We next demonstrate the QS for the case of a heterogeneous system where Eq. (\ref{Eq_8}) must be applied, in the form of surface stress of nanostructures, as in the case of epitaxial growth of nanometer-thick metal thin films when the quantum size effect is prominent. All the crystalline solid surfaces have a non-zero intrinsic surface stress \cite{ref1,ref14}. Classically, one may view this by imagining that the surface layer of atoms would have a smaller lattice constant (bond length) if they are separated by themselves; they have to be stretched apart when placed on top of the film to match the lattice constant of underlying film. For this reason, most solid surfaces have a tensile surface stress with a well-defined magnitude, a characteristic surface property of a given film structure. However, if the thickness of a film is reduced to nanoscale comparable to the de Broglie wavelength of electrons, quantum confinement becomes prominent giving rise to the formation of discrete quantum well states, known as the quantum size effect (QSE) \cite{ref15,ref16}. The QSE has been shown to modify surface energy \cite{ref17}. Here, we demonstrate that QSE will also modify surface stress, as a distinct manifestation of QS induced by quantum confinement \cite{ref18}.

Figure 3 shows the calculated surface energy ($\gamma$) and surface stress ($\sigma$) as a function of Pb(111) film thickness ($d$). $\gamma$ displays an oscillatory dependence on $d$, as known before \cite{ref17}.  What's new is that $\sigma$ displays a much stronger oscillatory dependence on $d$, showing a much stronger QSE on surface stress than on surface energy. The thickness dependence of the quantum surface stress can be understood from the thickness dependence of the quantum well states formed in the thin film, which modulates the thin film deformation potential ($\Xi$) and surface charge density ($\Delta n$) as a function of the film thickness, leading to the oscillating quantum stress. Because both $\Xi$ and $\Delta n$ are nonuniform in thin film, the simplified expression of Eq. (\ref{Eq_9}) cannot be used. The results in Fig. 3 are the integrated results of Eq. (\ref{Eq_8}) for each film thickness. Empirically, however, we may conveniently divide the surface energy into classical and quantum contributions as $\gamma=\gamma^C+\gamma^Q(d)$. Then, by definition, we express surface stress as
\begin{equation}
\label{Eq_10}
 \sigma=\frac{1}{A}\frac{d\gamma}{d\varepsilon}=\frac{1}{A}\frac{d\gamma^C}{d\varepsilon}+\frac{1}{A}\frac{d\gamma^Q}{d\varepsilon}=\sigma^C+\sigma^Q(d)
\end{equation}

where $A$is surface area, which is also divided into classical ($\sigma^C$) and quantum contribution ($\sigma^Q$). $\gamma^C$ and $\sigma^C$ represent, respectively, the classical surface energy (bond breaking energy) and surface stress (bond deformation energy) of a macroscopic thick film independent of film thickness; $\gamma^Q$ and $\sigma^Q$ represent, respectively, the quantum surface energy and surface stress, arising from quantum confinement in a nanoscale thin film, as a function of film thickness $d$. As the film thickness increases, $\gamma^Q$ and $\sigma^Q$ will eventually diminish and the system resumes the classical behavior.

\emph{Implications of QS}. The concept of QS has implications in a broad range of physical phenomena and technological applications that are based on coupling of electronic structure with lattice strain. We have already shown that the DFT calculation of QS provides an effective method for deriving deformation potential without the need of calculating band structure, which circumvents the difficulties encountered by previous methods as well as saves computational time. Physically, the QS induced by charge carriers will help us to better understand the physical nature of semiconductor doping in terms of the dopant-induced lattice stress, by differentiating the QS induced by electrons and holes from the CS induced by size difference between dopant and host atoms \cite{ref19}. In general, it is easier to dope an element whose QS and CS compensate each other, i.e, small n-type dopants or large p-type dopants, which induce the smallest overall amount of stress. There is a partial cancellation effect on the QS between the n- and p-type dopant, which makes the co-doping of both types of dopants easier.

\begin{figure}[b]
\includegraphics[width=3.4in]{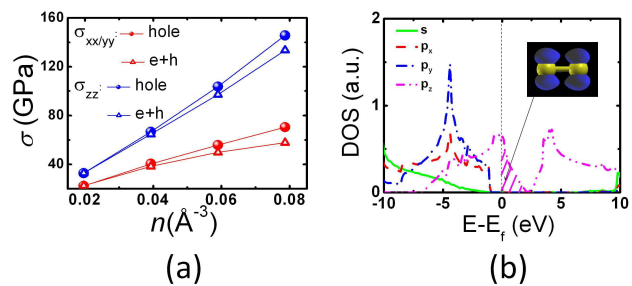}
\caption{(Color online) (a) Stress components (the z-component along the c-axis) induced by charge carriers as a function of carrier density in an ABC-stack graphite lattice. (b) The hole partial density of states with one electron removed from a 6-atom hexagonal graphite unit cell. The insets show the corresponding differential hole charge density of $p_z$ orbital shape. \label{fig4}}
\end{figure}

Another QS-related physical phenomenon is the pulse laser induced structural phase transition, such as the graphite-to-diamond transition \cite{ref20,ref21}. In a pulse lasing experiment, a high density of charge carriers (electrons, holes and excitons) is photo-excited in a small volume for a very short time. We argue that such charge carriers exert a large QS to the local lattice, causing effectively a "pressure-induced" structural phase transition. To support our point of view, we calculated the QS exerted by the photo-excited carriers (holes in the valence bands plus "free" electrons) to an ABC-stack or AB-stack graphite lattice \cite{suppl}. The QS is found to be tensile and highly anisotropic with the largest component along the z-axis and all the components increase approximately linearly with carrier density (Fig. 4a for the ABS-stack graphite). This is because the QS is dominated by the contribution from holes in the valence band of $p_z$ orbital, as indicated in Fig. 4b. The magnitude of the QS induced by a single hole in the 6-atom cell is as high as 20-30 GPa (Fig. 4a), which indicates that the pulse laser can induce a huge "local" stress (pressure) in the graphite lattice, larger than the critical pressure needed for the graphite-to-diamond transitions \cite{ref22}. Furthermore, we relaxed the graphite structure under the QS exerted by the charge carrier, we directly observed the lattice transformation of graphite into cubic diamond as the QS is gradually decreased upon structural optimization  \cite{suppl}. These results shed new lights on the understanding of the pulse laser induced graphite-to-diamond transition. More generally, we expect that the QS generated by the excited carrier is a key physical parameter in understanding a wide range of phase transitions induced by radiation of energetic particles\cite{Siamorp}.


The realization of the surface QS in nanoscale thin films has a profound effect on film stability and growth, especially when the film is strained. The QSE has been shown to be dominant in the growth of metal thin films on semiconductor substrates \cite{ref23}. On the other hand, the strain effect is ubiquitous in heteroepitaxial growth \cite{ref9,ref10}. So far, however, these two important effects have been usually studied separately focusing on one while neglecting the other, despite the fact they are both present in many cases. One outstanding difficulty is the lack of a theoretical framework underlying the fundamental relationship between the two effects. The introduction of the surface QS establishes a direct link between the quantum size and strain effects on the surface energy of thin films, underlying the interplay between the quantum size and strain effects on thin film stability and allowing a theoretical prediction of the thickness-dependent surface energy in a strained quantum film.

In conclusion, we present a rigorous derivation of QS within DFT, which is shown in linear relation to deformation potential and charge density variation, as a quantum analog of Hook's law of CS. Two distinct physical manifestations of QS induced by charge carriers and quantum confinement are demonstrated by direct DFT calculations, confirming the theoretical formulation. The concept of QS adds an important quantum mechanical aspect to our classical view of stress. It manifests itself broadly in various mechanical properties of different material systems, especially in low-dimensional nanostructures where quantum effects become prominent \cite{ref18}. It will be especially relevant in advancing our fundamental understanding of physical phenomena and technological applications that are based on coupling of electronic structure with lattice stress, such as charge carriers confined in quantum wells, wires and dots, free carriers created in nanofilms by electrical gating, electroelastic effects, magnetoelastic effects and biological cell deformation due to charging and polarization.

This work was support by DOE-BES program (Grant No. DE-FG02-04ER46148). In addition, M. Liu thanks support from NSF-MWN and Materials Theory program (Grant No. DMR-0909212) and H. Ding thanks support from DOE-EFRC (Grant No. DE-SC0001061). We thank also DOE-NERSC and the CHPC at University of Utah for providing the computing resources.




\begin{thebibliography}{99}
\bibitem{ref1}  R. C. Cammarata,Progress in Surf. Sci. 46, 1 (1994).
\bibitem{ref2}  M. Ieong {\em et al}., Science 306, 2057 (2004).
\bibitem{ref3}  F. Sch$\ddot{a}$ffler, Semicond. Sci. Technol. 12, 1515 (1997).
\bibitem{ref4}  V. Sazonova, Nature 31,284 (2004).
\bibitem{ref5}  J. Zang and F. Liu, Nanotechnology 18, 405501 (2007).
\bibitem{ref6}  D.-Y. Khang {\em et al}., Science 311, 208 (2006).
\bibitem{ref7}  M. Huang {\em et al}., Adv. Mat. 17, 2860 (2005).
\bibitem{ref8}  D. Yu and F. Liu, Nano Lett. 7, 3046 (2007).
\bibitem{ref9}  F. Liu and M. G. Lagally, Surf. Sci. 386, 169 (1997).
\bibitem{ref10} F. Liu, Phys. Rev. Lett. 89, 246105 (2002).
\bibitem{ref11} P. Hohenberg and W. Kohn, Phys. Rev. 136, B864 (1964).
\bibitem{suppl} See Supplemental Material for details of DFT QS calculations and discussions of the photocarrier induced graphite-to-diamond transition.
\bibitem{ref12} O. H. Nielsen and R. M. Martin, Phys. Rev. Lett. 50, 697 (1985).
\bibitem{ref13} Landolt-B\"{o}rnstein, Numerical Data and Functional Relationships in Science and  Technology, Group III, Vol. 22a, edited by O. Madelung and M. Schulz (Springer, Berlin, 1987).
\bibitem{ref14} F. Liu and M. G. Lagally, Phys. Rev. Lett. 76, 3156(1996).
\bibitem{ref15} F. K. Schulte, Surf. Sci. 55. 427(1976).
\bibitem{ref16} F. Liu, F. N. Khanna, and P. Jena, Phys. Rev. B 42, 976 (1990).
\bibitem{ref17} C. M. Wei and M. Y. Chou, Phys. Rev. B  66, 233408 (2002).
\bibitem{ref18} B. Huang et al., Phys. Rev. Lett. 102, 166404 (2009).
\bibitem{ref19} J. Zhu et al., Phys. Rev. Lett. 105, 195503 (2010).
\bibitem{ref20} R. K. Raman, Phys. Rev. Lett. 101, 077401(2008).
\bibitem{ref21} J. Kanasaki et al., Phys. Rev. Lett. 102, 087402 (2009).
\bibitem{ref22} R. Clarke and C. Uher, Adv. Phys. 33, 469 (1984).
\bibitem{Siamorp} M. Harb et al., Phys. Rev. Lett. 80, 5381(1998).
\bibitem{ref23} Z. Zhang, Q. Niu, and C. K. Shih, Phys. Rev. Lett. 80, 5381(1998).

\end{thebibliography}
\end{document}